\documentclass[10pt,conference]{IEEEtran}

\newcommand{\add}[1]{{\color{mygreen}{#1}}}
\newcommand{\sub}[1]{{\color{red}{#1}}}
\newcommand{\myComment}[1]{{\color{mygray}{#1}} \par}
\newcommand{\review}[1]{{\color{mygray}{#1}} \par}

\renewcommand{\sub}[1]{}
\renewcommand{\myComment}[1]{}
\renewcommand{\add}[1]{#1}
\renewcommand{\review}[1]{}

\usepackage{common}

\begin{document}

\title{\irfuzzer{}: Specialized Fuzzing for LLVM Backend Code Generation}
\makeatletter
\newcommand{\linebreakand}{%
  \end{@IEEEauthorhalign}
  \hfill\mbox{}\par
  \mbox{}\hfill\begin{@IEEEauthorhalign}
}
\makeatother
\author{
    \IEEEauthorblockN{Yuyang Rong\IEEEauthorrefmark{1}\IEEEauthorrefmark{2}}
    \IEEEauthorblockA{
        PeterRong96@gmail.com
    }
    \and
    \IEEEauthorblockN{Zhanghan Yu\IEEEauthorrefmark{2}}
    \IEEEauthorblockA{
        hnryu@ucdavis.edu
    }
    \and
    \IEEEauthorblockN{Zhenkai Weng\IEEEauthorrefmark{2}}
    \IEEEauthorblockA{
        zweng@ucdavis.edu
    }
    \and
    \IEEEauthorblockN{Stephen Neuendorffer\IEEEauthorrefmark{1}}
    \IEEEauthorblockA{
        stephen.neuendorffer@amd.com
    }
    \and
    \IEEEauthorblockN{Hao Chen\IEEEauthorrefmark{2}}
    \IEEEauthorblockA{
        chen@ucdavis.edu
    }
    \linebreakand
    \IEEEauthorblockA{
        \IEEEauthorrefmark{1}Advanced Micro Devices, Inc. \\
        \IEEEauthorrefmark{2}University of California, Davis \\
    }
}

\maketitle
\begin{abstract}
    Modern compilers, such as LLVM, are complex.
    Due to their complexity, manual testing is unlikely to suffice, yet formal verification is difficult to scale.
    End-to-end fuzzing can be used, but it has difficulties in discovering LLVM backend problems for two reasons.
    First, frontend preprocessing and middle optimization shield the backend from seeing diverse inputs.
    Second, branch coverage cannot provide effective feedback as LLVM backend contains much reusable code.
    In this paper, we implement \irfuzzer{} to investigate the need of specialized fuzzing of the LLVM compiler backend.
    We focus on two approaches to improve the fuzzer: guaranteed input validity using constrained mutations to improve input diversity and new metrics to improve feedback quality.
    The mutator in \irfuzzer{} can generate a wide range of LLVM IR inputs, including structured control flow, vector types, and function definitions.
    The system instruments coding patterns in the compiler  to monitor the execution status of instruction selection.
    The instrumentation not only provides new coverage feedback on the matcher table but also guides the mutator on architecture-specific intrinsics.

    We ran \irfuzzer on \numCPU{}  mature LLVM backend targets. \irfuzzer discovered \numTotalBugs{} new, confirmed bugs in LLVM upstream, none of which existing fuzzers could discover. This demonstrates that \irfuzzer is far more effective than existing fuzzers. Upon receiving our bug report, 
    the developers have fixed \numFixedBugs{} bugs and back-ported \backPortFix{} fixes to LLVM 15, which shows that specialized fuzzing provides actionable insights to LLVM developers.
\end{abstract}


\section{Introduction}
\label{sec:introduction}


Modern compilers, such as LLVM~\cite{lattner2004llvm}, are complex software.
For example, LLVM consists of over seven million lines of C/C++ code contributed by more than \SI{2500}\xspace developers. 
Given the size of this codebase and its importance in the computing ecosystem, an effective, scalable verification method is critical.
Despite extensive testing, latent bugs remain and their impact on users can be quite significant given the widespread distribution and long lifetime of compilers.

To reduce latent bugs, various techniques have been used to automate the verification of compilers, such as partial model checking~\cite{lopes2021alive2}, fuzzing~\cite{LibAFL, CSmith, 10.1145/3428264}, and differential testing~\cite{emi,spe}.
Although end-to-end formal verification of compilers has been applied \cite{Jourdan-LBLP-2015, Bourke-BDLPR-2017}, these techniques have not yet scaled to practical compilers such as LLVM, which supports a wide range of architectures, programming languages, and use models. 

In the specific case of LLVM, another factor making verification difficult is that the interface between compiler optimization and machine code generation is widely used but not completely specified.
As a result, it can be difficult for backend developers to understand whether they have completely implemented the wide range of possible inputs.
In addition, backends often differ greatly in their relative code maturity, including some targets that are relatively mature and other targets for new devices that are in active development.

We find that the \sota fuzzers failed to find new bugs of a compiler backend for various reasons.
General-purpose fuzzing techniques, such as AFL++~\cite{AFLplusplus}, often do not consider input validity and struggle to explore control paths in the compiler backend since most binary strings are invalid compiler inputs.
In order to test the compiler backend more effectively, we aim to generate LLVM Intermediate Representation (LLVM IR) that complies with the language reference.
LLVM includes  \texttt{llvm-opt-fuzzer} and \texttt{llvm-isel-fuzzer}, which generate valid IR for middle end and backend fuzzing, respectively~\cite{LLVMFuzzing}.
Both of them are based on the library \fuzzmutate{}~\cite{FuzzMutate} for valid IR mutation.
However, \fuzzmutate{} can't construct complex control flows and only generates a few instructions with scalar types.
On the other hand, end-to-end fuzzing tools, such as CSmith~\cite{CSmith} and GrayC~\cite{GreyC}, test the whole pipeline of the compiler, but they cannot explore control paths in the compiler backend efficiently.
CSmith does not take any feedback from the compiler, which contributes to its ineffectiveness.
A more fundamental reason is that front-end parser and middle-end optimizations may limit the set of features seen by the compiler backend.
High level languages, such as C, may not exercise all backend features in LLVM.
Therefore, even if GrayC used branch coverage feedback from libFuzzer~\cite{serebryany2016continuous}, it missed many backend bugs introduced before LLVM 12, which were found by us.
As a result, when a new language, such as Rust, is introduced, new backend bugs may still arise~\cite{rustbug}.

Generating valid IR is challenging with three major difficulties.
In order to generate a complex control flow graph (CFG), we have to maintain all data dependencies to avoid use-before-definition situations.
A valid CFG can be easily invalidated by a jump, as shown in \autoref{fig:cfg}.
This challenge does not exist in C generation as long as one does not generate \texttt{goto} statements.
Besides, modelling the instructions missing in \fuzzmutate{} isn't trivial.
We must make sure that the types of the operands in each \ir{} match, but enumerating the large numbers of natively supported vector types is infeasible.
Finally, it is difficult to model intrinsic functions for all architectures, as intrinsics are often poorly documented and vary from architecture to architecture.

We also observe that AFL++'s feedback mechanism performed poorly when testing the backend.
It uses branch coverage as feedback, which runs into severe branch collision problems when fuzzing large code bases such as LLVM.
Naively increasing the branch counting table size introduces huge overhead ~\cite{8418631}.
A more fundamental reason is that much code generation logic in the LLVM backend is implemented using table-driven state machines.
A matcher table encapsulates all possible states as a constant byte array, meaning that branch counting can't observe this logic during fuzzing.
The fuzzer needs a better feedback on whether the seed is interesting or not.
If the seed is not interesting, the feedback should also inform the mutator what type of input is desired.

\begin{figure*}[t]
    \centering
    \includegraphics[width=\linewidth]{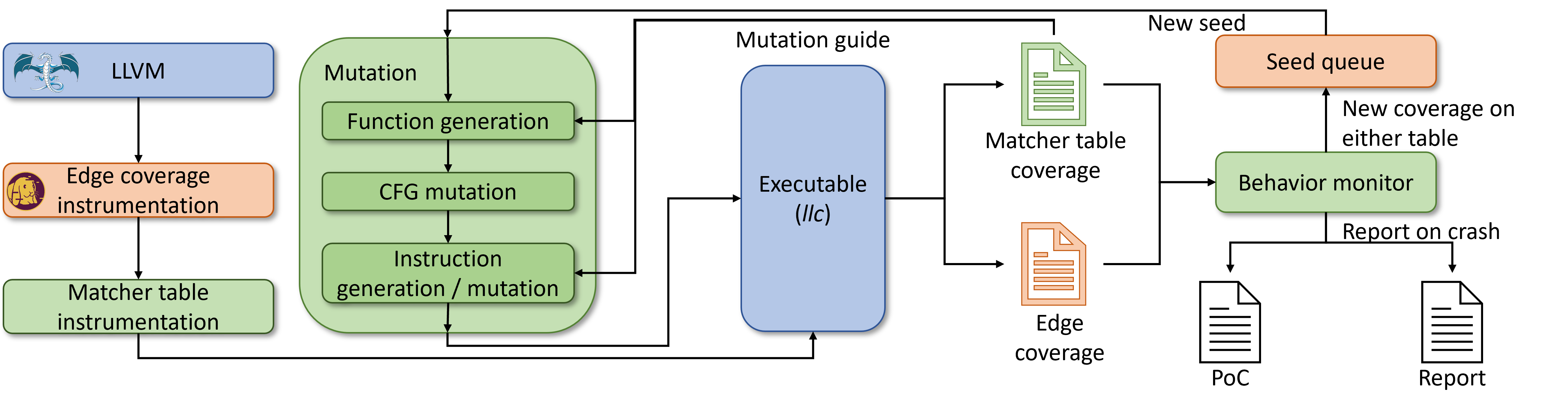}
    \caption{
        Overview of \irfuzzer.
        Green shaded components are the contributions of this paper, orange shaded components are AFL++, and blue shaded components are from LLVM.
        We created an LLVM IR mutator that guarantees the correctness of the generated input (\autoref{sec:design:mutation}).
        We introduced a new coverage metric to track the backend code generation while guiding the mutation module (\autoref{sec:design:coverage}).
    }
    \vspace{-\baselineskip}
    \label{fig:overview}
\end{figure*}

\begin{figure*}[t]
    \centering
    \begin{subfigure}[t]{\linewidth}
        \centering
        \begin{tikzpicture}[auto]
            \tikzstyle{every node} = [align=center, node distance=2.5cm]
            \node [draw, circle] (B1)  {B1};
            \node [draw, circle, right of=B1] (B2) {B2};
            \node [draw, circle, right of=B2] (B3) {B3};

            \path (B1) edge[cfgedge] node {} (B2);
            \path (B2) edge[cfgedge] node {} (B3);
        \end{tikzpicture}
        \caption{Original CFG}
        \label{fig:cfg:a}
    \end{subfigure}

    \begin{subfigure}[t]{0.44\linewidth}
        \centering
        \begin{tikzpicture}[auto]
            \tikzstyle{every node} = [align=center, node distance=2.5cm]
            \node [draw, circle] (B1)  {B1};
            \node [draw, circle, right of=B1] (B2) {B2};
            \node [draw, circle, right of=B2] (B3) {B3};

            \path (B1) edge[cfgedge] node {} (B2);
            \path (B2) edge[cfgedge] node {} (B3);
            \path (B1) edge[dashedcfgedge, bend left] node {} (B3);
        \end{tikzpicture}
        \caption{Incorrect mutation: randomly adding an edge}
        \label{fig:cfg:b}
    \end{subfigure}
    \hfill
    \begin{subfigure}[t]{0.55\linewidth}
        \centering
        \begin{tikzpicture}[auto]
            \tikzstyle{every node} = [align=center, node distance=1.5cm]

            \node [draw, circle] (B1)  {$B1$};
            \node [draw, circle, right of=B1] (BS) {$B2$};
            \node [draw=none, above=0.1mm of BS] (src) {\textit{entry}};
            \node [draw, dashed, circle, right of=BS, minimum size=0.8cm] (sCFG) {$\cdots$};
            \node [draw=none, above=0.1mm of sCFG] (sCFGText) {\sCFG};
            \node [draw, circle, right of=sCFG] (BT) {$B2$};
            \node [draw=none, above=0.1mm of BT] (sink) {\textit{exit}};
            \node [draw, circle, right of=BT] (B3) {$B3$};

            \path (B1) edge[cfgedge] node {} (BS);
            \path (BS) edge[dashedcfgedge] node {} (sCFG);
            \path (sCFG) edge[dashedcfgedge] node {} (BT);
            \path (BT) edge[cfgedge] node {} (B3);
        \end{tikzpicture}
        \caption{Correct mutation: splitting node and adding \sCFG}
        \label{fig:cfg:c}
    \end{subfigure}
    \caption{
        Examples of incorrect and correct CFG mutations.
        \autoref{fig:cfg:a} is the original linear CFG.
        \autoref{fig:cfg:b} naively adds an edge from B1 to B3. After that, B2 no longer dominates B3, so every value defined in B2 and used in B3 may cause use-before-definition error.
        \autoref{fig:cfg:c} breaks B2 into an entry and an exit node and inserts an sCFG between them. This guarantees that B2 still dominates B3.
    }
    \vspace{-\baselineskip}
    \label{fig:cfg}
\end{figure*}

To address these issues, we design a specialized fuzzer, \irfuzzer{}, for fuzzing the LLVM compiler backend.
\autoref{fig:overview} shows the overall structure of \irfuzzer.
We first design a mutator that generates valid IR (\autoref{sec:design:mutation}).
We maintain the domination relation in a CFG during mutation by inserting subgraphs (\sCFG) into the existing CFG as shown in ~\autoref{fig:cfg:c}.
We also use a descriptive language to list the requirements of each instruction type.
This approach ensures that inputs to the compiler backend are \textit{always} valid, increasing the efficiency of fuzzing.
Our work expands \fuzzmutate{} to include special handling by compiler backends, such as multiple basic blocks with complex control flow, function calls, intrinsic functions, and vector types.
Using \irfuzzer{}, we are able to generate a wider range of instructions and explore control paths in the compiler backends more efficiently.

Then, we introduce a new coverage metric (\autoref{sec:design:coverage}) by instrumenting the table-driven state machines in LLVM, enabling the design space to be more efficiently explored.
New entries covered in the matcher table indicate that new features are executed.
Working together with branch coverage, they provide better feedback on whether a seed is interesting.
Furthermore, the matcher table contains all the information about the instructions and intrinsics in one architecture.
As a result, we use the matcher table to determine which instructions and intrinsics haven't been fuzzed.
We design a feedback loop from the matcher table coverage to our mutator.
\irfuzzer{} periodically sends to the mutator a coverage report containing the states that haven't been executed to guide mutations. This allows \irfuzzer{} to test on different backends with no prior knowledge of the architecture.

We evaluated \irfuzzer{} on \numCPU{} mature backend architectures in LLVM (\autoref{sec:eval}).
Our results show that \irfuzzer{} is more effective than the \sota fuzzers AFL++ and GrayC.
\irfuzzer{} generated inputs code with better branch coverage and matcher table coverage on \sub{28} \add{all} LLVM backends.
Leveraging these techniques, we were able to find and report \numTotalBugs{} new bugs in LLVM, of which all have been confirmed, \numFixedBugs{} have been fixed, and \backPortFix{} have been back ported to LLVM 15.
This demonstrates the high impact on improving the correctness of LLVM backend targets.

This paper uses LLVM to demonstrate the importance of having a specialized fuzzer for the compiler backend.
Since modern compilers have similar intermediate representations, we expect that our approach can apply to other compilers without requiring heavy engineering efforts.
We made the following contributions in this paper:
\begin{itemize}
    \item We designed and implemented \irfuzzer{}. To the best of our knowledge, \irfuzzer{} is the first backend fuzzer that uses matcher table coverage feedback to guide mutation.
    \item We compared \irfuzzer{} with other \sota{} fuzzers on LLVM upstream and found it to be the most effective on matcher table coverage.
    \item  We carefully analyzed and categorized the bugs we found during our testing. In total, we discovered \numTotalBugs{} confirmed new bugs in LLVM, of which \numFixedBugs{} have been fixed and \backPortFix{} have been back ported to LLVM 15.
\end{itemize}

\section{Background}
\label{sec:background}

\subsection{LLVM}
\label{sec:background:LLVM}

LLVM~\cite{lattner2004llvm} is a mature compiler framework consisting of many components that can be targeted to different architectures.
At its core lies the LLVM Intermediate Representation (LLVM IR), which serves as a target-independent abstraction separating the concerns of high-level programming languages from the low-level details of particular architectures.
LLVM can be roughly partitioned into three layers. 
The \textit{frontend}, such as \texttt{clang}, translates programming languages to LLVM IR, including lexer, parser, AST transformation, etc.
The \textit{middle-end}, called \texttt{opt}, processes LLVM IR and performs code analysis and many common target-independent optimizations.
The \textit{backend}, called \texttt{llc}, converts LLVM IR to a target-specific machine code representation and eventually assembly code for the target architecture. 
The LLVM backend supports multiple target architectures through a plug-in abstraction, and the code to support a target architecture typically involves the implementation of API functions to describe common aspects along with target specific code to implement more unusual concepts.

The LLVM IR describes a static single-assignment (SSA) form~\cite{SSA}, with a fixed set of instructions.
Instructions are strongly typed, and the type of each value must match between its definition and all uses.
A wide range of types are supported, including integers with arbitrary bitwidth, floating point values, pointers, vectors, and other aggregate types.
As with most high-level languages, LLVM IR allows the definition of functions, and the control flow between functions is implemented using the \texttt{call} instruction. Architecture specific intrinsics have no corresponding IR instructions, but are represented as function calls at IR level.

Control flow within a function in LLVM IR is represented using basic blocks and branch instructions.
Special PHI instructions allow instructions in a basic block to refer to values defined in other basic blocks.
Therefore, PHI instructions must respect control flow constraints and may only refer to values defined in predecessor blocks.
This \textit{domination constraint}~\cite{prosser1959applications} means that techniques used in high-level language generation cannot be easily adapted to LLVM IR.



The process of \textit{instruction selection} in the LLVM backend replaces target-independent LLVM IR instructions with target-specific machine code instructions.
LLVM provides two different frameworks to implement instruction selection that may be leveraged by the target backend plug-in.
\dagisel{} ~\cite{SelectionDAG} is the more mature instruction selection framework and is leveraged by all targets.
In \dagisel{}, the code in each basic block is converted into a directed acyclic graph (DAG) representing the data dependency between instructions, and instruction selection is performed on the DAG.
Since \dagisel{} processes each basic block independently, it can miss opportunities for optimization across basic blocks.
\gisel{}~\cite{GlobalIsel} is a newer framework that is only leveraged by some targets.
\gisel{} preserves the basic block structure within a function during instruction selection, enabling more optimization opportunities.
\sub{
    Both frameworks use \textit{patterns} to describe rewrite rules applied during instruction selection.
    Some patterns are relatively simple and replace a single LLVM IR instruction with a single machine instruction.
    More complicated patterns may replace multiple LLVM IR instructions, or generate multiple machine instructions.
    Patterns may also include complex predicates to limit their applicability only to specific situations.  For example, a pattern may only apply when a particular operand is a constant, or a certain hardware feature is enabled.


    Most patterns are described declaratively in an LLVM-specific language called TableGen~\cite{TableGen}.
    In order to optimize the application of patterns, TableGen translates individual patterns into a state-machine representation implemented as a large byte array in C++ known as the \textit{matcher table}.
    During compilation, the state machine in the matcher table is executed on each \ir{} and determines the best pattern to apply.
    \autoref{lst:codegen} is a C++ code snippet used to evaluate the matcher table in \dagisel{}.
    \texttt{SDNode} is a data structure that represents an \ir{}.
    The while loop iteratively reads a command from the matcher table based on the current state, represented by the \texttt{idx} variable, evaluates the command, and selects the next state that will be evaluated.
    For example, \texttt{Opc\_CheckOpcode} will check if the opcode of a given \texttt{SDNode} representing an instruction in the \dagisel{} graph matches a particular opcode.
    The \texttt{Result} will be used in future iterations, depending on the next entry in the matcher table.
    Evaluation of the matcher table continues until a single pattern is selected, or a state is reached where no patterns can apply.

    Since the program in \autoref{lst:codegen} evaluates all the patterns using the same set of conditional branches in the switch statement, its control flow coverage cannot measure what patterns have been exercised.
}



\subsection{Coverage guided fuzzing}
\label{sec:background:fuzzing}

American Fuzzy Lop (AFL)~\cite{AFL}, an open source fuzzing framework, implements coverage-guided fuzzing. 
It instruments the program under test (PUT) with the ability to track control-flow coverage.
When an input increases code coverage, AFL stores it in a \textit{seed cache} and mutates it to generate new inputs. This strategy allows AFL to explore different control-flow paths of the PUT efficiently.

Many variations of coverage guided fuzzing have been developed, with the goal of finding bugs more efficiently by exploring a wider range of program behaviors with future executions of the PUT~\cite{10.1145/3512345}.
There are studies on the impact of different feedback algorithms~\cite{241986, osti_10313742, IJON}.
Different methods are proposed to prioritize seeds to improve the performance of fuzzing.~\cite{aflfast, MOPT, she2022effective}.
Some fuzzers target specific bugs~\cite{251544,rong2020int, razzer} and libraries~\cite{chen2023Hopper, lyu2024PromptFuzz}.
More advanced mutation strategies also show better fuzzing performance compared with random mutation~\cite{redqueen, neuzz, angora, chen2019Matryoshka, 10.1145/3510003.3510174, rong2022valkyrie}.
Many improvements have been implemented in AFL++~\cite{AFLplusplus}, making it a good framework for further development. Fuzzing not only finds bugs but also helps program understanding~\cite{zhao2023}.

LLVM also introduces its own coverage guided fuzzing framework libFuzzer~\cite{serebryany2016continuous}, coupled with \fuzzmutate{}~\cite{FuzzMutate}, it can be used to fuzz LLVM backend.
However, \fuzzmutate{} only generates a limited type of code and is not under active development.
Still, the framework provides us with helpful insights into how should we mutate LLVM IR.

\section{Design}
\label{sec:design}

\begin{figure}[t]
    \begin{subfigure}{\linewidth}
        \begin{lstlisting}[
    language=llvm, 
    caption={
        A piece of LLVM IR program generated by function generation (\autoref{sec:design:mutation:func}). 
        The function returns a 64 bit integer, so we allocate a stack memory and load from it to return. We will fill the memory in later mutations.
    }, 
    label=lst:ir:a,
    basicstyle=\ttfamily\scriptsize,
]
define i64 @f(i32 zeroext %I, <4 x i32> %V) noinline  { 
Entry:    %ret_p = alloc i64, 1
          %ret = load i64, ptr %ret_p
          ret i64 %ret
}
        \end{lstlisting}
    \end{subfigure}
    \begin{subfigure}{\linewidth}
        \begin{lstlisting}[
            language=llvm, 
            caption={
                IR program mutated from \autoref{lst:ir:a}. 
                Line \autoref{lst:ir:b:begin} to \autoref{lst:ir:b:end} are introduced by \sCFG insertion (\autoref{sec:design:mutation:cfg}). 
                We insert \sCFG by splitting the \texttt{Entry} block into two and generating a \texttt{switch} instruction.
            }, 
            label=lst:ir:b,
            basicstyle=\ttfamily\scriptsize,
        ]
define i64 @f(i32 zeroext %I, <4 x i32> %V) noinline { 
EntrySrc:     %ret_p = alloc i64, 1
              switch i32 %I, label %sCFG_Default [  (*@ \label{lst:ir:b:begin} @*)
                i32 1, label %sCFG_1
              ]
sCFG_Default: br label %EntrySink
sCFG_1:       br label %EntrySink (*@ \label{lst:ir:b:end} @*)
EntrySink:    %ret = load i64, ptr %ret_p
              ret i64 %ret
}
        \end{lstlisting}
    \end{subfigure}
    \begin{subfigure}{\linewidth}
        \begin{lstlisting}[
            language=llvm, 
            caption={
                IR program mutated from \autoref{lst:ir:b}.
                Instruction insertion(\autoref{sec:design:mutation:inst_gen}) generates line \autoref{lst:ir:c:zext}, \autoref{lst:ir:c:call}, and \autoref{lst:ir:c:phi}. 
                The placeholder memory is also used by \texttt{\%PHI} to avoid undefined behavior (Line \autoref{lst:ir:c:store}).
            }, 
            label=lst:ir:c,
            basicstyle=\ttfamily\scriptsize,
        ]
define i64 @f(i32 zeroext %I, <4 x i32> %V) noinline { 
EntrySrc:    %ret_p = alloc i64, 1
             switch i32 %I, label %sCFG_Default [
               i32 1, label %sCFG_1
             ]
sCFG_Default:%I64 = zext i32 %I, i64 (*@ \label{lst:ir:c:zext} @*)
             br label %EntrySink
sCFG_1:      %I1 = add i32 %I, 1 (*@ \label{lst:ir:b:2} @*)
             %J64 = call @f(i32 %I1, <4 x i32> %V) (*@ \label{lst:ir:c:call} @*)
             br label %EntrySink
EntrySink:   %PHI = phi i64 [%J64, %sCFG_1], 
                            [%I64, %sCFG_Default](*@ \label{lst:ir:c:phi} @*)
             store i64 %PHI, %ret_p (*@ \label{lst:ir:c:store} @*)
             %ret = load i64, ptr %ret_p
             ret i64 %ret
}
        \end{lstlisting}
    \end{subfigure}
    \caption{
        An example of how \irfuzzer{} mutates a module using different strategies.
    }
    \label{fig:ir}
\end{figure}
We design \irfuzzer{} with two new components.
\autoref{fig:overview} shows the new components of \irfuzzer{}.
During the mutation stage, we first generate a function if there isn't one (\autoref{sec:design:mutation:func}).
Then we change the control flow graph (CFG) to create more control flows (\autoref{sec:design:mutation:cfg}).
Finally, we generate new \ir{s} and mutate them (\autoref{sec:design:mutation:inst_gen}).
\sub{
    \autoref{fig:ir} shows an example of the mutation process using these mutation strategies.
}
After the mutation stage, we create a new method to measure the coverage of the program (\autoref{sec:design:coverage}).
Although \irfuzzer borrows from FuzzMutate and other tools, all the components described in this section are novel unless otherwise noted.

\subsection{LLVM IR mutation}
\label{sec:design:mutation}

To generate a wide variety of input while avoiding invalid inputs, we adopt a mutation-based strategy.
This strategy starts with small valid seed inputs and modifies the seed inputs in ways that should also generate valid inputs.
By randomly selecting between a number of small, well-defined mutations, we expect to eventually generate a broad class of valid inputs while avoiding invalid inputs.
\autoref{fig:ir} shows an example of our mutator in practice.
We first generate an empty function if none is present (\autoref{lst:ir:a}).
Then, we mutate the control flow by \sCFG insertion (\autoref{lst:ir:b}).
Finally, we modify or insert instructions in basic blocks (\autoref{lst:ir:c}).


\subsubsection{Function generation}
\label{sec:design:mutation:func}

The LLVM backend has many target-specific code related to function calls.
As a result, it is important to generate a wide range of function definitions and function calls with different arguments and return types.

\irfuzzer{} implements a mutation strategy capable of generating new function definitions with arbitrary arguments and return types.
One important constraint is that the return type of the function signature matches the type of each \texttt{return} instruction in the function definition.
To ensure this, \irfuzzer synthesizes a load instruction of an appropriate type as the operand for a \texttt{return} instruction.
Although the value returned from the load may be uninitialized, later mutations may store values to the memory, validating the return value.

\irfuzzer{} also implements a mutation strategy to generate new \texttt{call} instructions that refer to specific function declarations.
The mutator is free to select from any declared functions and generate compatible arguments and return values for the call, as with any other primitive instruction.
Intrinsic functions are target specific operations that correspond to complicated \mir{s}, so generating them will increase the code we can test.
Yet they are treated as functions at middle-end.
In particular, this mutation strategy will also select intrinsic functions to call.

\myComment{Addressing meta review 4.}
\add{
    Function attributes can impact backend behavior.
    These attributes are often set by the compiler frontend and middle end to optimize the code.
    For example, \texttt{noinline} can change how a function is treated during code generation.
    LLVM categorizes attributes into function attributes, argument attributes, and return value attributes.
    To demonstrate \irfuzzer{}'s potential, we include 15 attributes in \irfuzzer{}.
    Most attributes do not affect the validity of the program.
    However, we need to model the contradicting ones to ensure that they do not appear at the same time, such as \texttt{OptForFuzzing} and \texttt{OptimizeForSize}.
}

\subsubsection{CFG mutation}
\label{sec:design:mutation:cfg}


\sub{
    Another area where target-specific code in the LLVM backend differs relates to control flow.
}\add{
    Generating diverse CFGs is necessary to test LLVM backends.
}
Many machine code optimizations\sub{, such as jump threading,} restructure control flow.
In addition, certain compiler optimizations may select specific jump instructions, but this optimization can only be performed after instruction selection when the code size and alignment are known.
For instance, a common compiler optimization is to first select jump instructions in a ``short'' form with a limited offset range and then only replace the short form with a ``long'' jump instruction if a larger offset is required.
\sub{Control flow optimization can also affect register lifetimes, exercising target-specific code for spilling and restoring values from the stack.}


\sub{
    \irfuzzer{} implements a structured approach to generate control flow.
    Inserting and removing arbitrary branches in the code can greatly change dominator constraints between basic blocks.
    For example, in \autoref{fig:cfg:b}, mutated from \autoref{fig:cfg:a}, $B2$ no longer dominates $B3$ after mutation.
    If any value in $B3$ refers to values in $B2$, the module is invalid after mutation.
    We implement an elegant approach that uses sub-control flow graph insertion, or \sCFG insertion, as shown in \autoref{fig:cfg:c}.
    Instead of changing edges, we split a block and insert \sCFG inside.

    A \sCFG is a CFG with a single \textit{source} entry block and a single \textit{sink} exit block that will be placed inside a larger CFG.
    Within the \sCFG, we allow the synthesis of an arbitrary control flow graph.
    However, every control flow edge starting in the \sCFG must be contained within the \sCFG, except for \textit{source} edges, \textit{sink} edges, and return instructions.
    With this restriction, we can insert \sCFG into a program without breaking the dominator constraint by randomly selecting a block and splitting it into two.
    The first part of the block is the \textit{source} and the second is the \textit{sink}.
    After block splitting, we generate random \sCFG starting from \textit{source} and ending with \textit{sink}.
}\add {
    A challenge in mutating CFGs is how to preserve the dominator constraints. Naively inserting and removing arbitrary edges in a CFG may violate dominator constraints between basic blocks, which may cause use-before-definition.
    For example, starting from the CFG in \autoref{fig:cfg:a}, after we add an edge, $B2$ no longer dominates $B3$ in \autoref{fig:cfg:b}. If $B3$ uses any value defined in $B2$, the program will cause a use-before-definition error if it follows the new edge.

    To overcome this problem, \irfuzzer{} inserts \emph{sub-control flow graphs} (\sCFG), as shown in \autoref{fig:cfg:c}, to maintain dominator constraints.
    \begin{definition}
    A sub-control graph (\sCFG) is a CFG with a single entry block and a single exit block. The exit block must have a single outgoing edge, and all the other blocks must either branch to another block in this \sCFG or \texttt{return}. 
    \end{definition}
}

\sub{
    The \sCFG can be constructed with three main control schemas: \texttt{branch}, \texttt{switch}, and \texttt{return}, corresponding to different terminators of the basic block.
    \irfuzzer{} starts with only one basic block and randomly selects the terminator of the block.
    If the return statement is selected and the function requires a return type, we pick any value available that matches the return type of the function.
    If a branch or switch statement is selected as the terminator, we find a previously generated non-constant value as a condition.
    If no such values can be found, we allocate a stack memory as a placeholder.
    The branch can go to one of three places: \textit{sink}, self-loop or return.
    If we generate a self-loop, we also update all the PHI nodes in the block to include a new value.
    Finally, all terminators are generated during CFG mutation, and instruction generation strategies will not mutate terminators to protect the integrity of the CFG.
}
\myComment{Addressing to meta review 2: ``Provide the correctness proof of the new (\sCFG-based) mutation operators.''}
\add{
    \begin{definition}
        A block $S$ dominates a block $T$ if $S$ precedes $T$ on all the reachable paths ending in $T$.
    \end{definition}
    \begin{theorem}
        Let block $S$ dominate block $T$. Let $B$ be a block, and $C$ be an sCFG. Then, after we split $B$ into $B_\text{entry}$ and $B_\text{exit}$ and insert $C$ between them, $S$ still dominates $T$.
        \label{thm:scfg}
    \end{theorem}
    
    \begin{proof}
        Before we insert $C$, since $S$ dominates $T$, $S$ precedes $T$ on every reachable path ending in $T$. Let $p = (\cdots, S, \cdots, T)$ be such a path. After we insert $C$,
        \begin{itemize}
            \item If no block in $C$ is inserted between $S$ and $T$ on $p$, then $p$ does not violate the property that $S$ dominates $T$.
            \item If some blocks in $C$ are inserted between $S$ and $T$ on $p$, then $p$ becomes one or more new paths $p' = (\cdots, S, \cdots, B_\text{entry}, C_\text{entry}, C*, C_\text{exit}, B_\text{exit},  T)$ where $C*$ represents a sequence of blocks in $C$. If no block in $C*$ contains a \texttt{return}, then $S$ still precedes $T$ on $p'$. On the other hand, if any block in $C*$ contains a return, then $T$ is not reachable on $p'$. In both these cases, $p'$ does not violate the property that $S$ dominates $T$.
        \end{itemize}
    \end{proof}

    \autoref{thm:scfg} guarantees that splitting a block and inserting an sCFG between them preserves dominator constraints. Based on this property, \irfuzzer mutates CFG in the following steps.
    \begin{enumerate}
        \item  \irfuzzer selects a block, and a non-terminating instruction inside it as the split point.

        \item \irfuzzer splits the block into an \emph{entry} block, containing all the instructions before the split point, and an  \emph{exit} block, containing all the remaining instructions. Then, it randomly generates a \texttt{branch} or \texttt{switch} instruction as the entry block's new terminating instruction.

        \item \irfuzzer creates empty blocks as the destinations of the \texttt{branch} or \texttt{switch} instruction created in the previously step. For each empty block, \irfuzzer randomly selects \texttt{branch}, \texttt{switch}, or \texttt{return} as its terminator.

        \item If \irfuzzer selects \texttt{branch} or \texttt{switch}, it either routes the control flow to the \textit{exit} block or creates a self loop.

    \end{enumerate}
}

\subsubsection{Instruction modeling and generation}
\label{sec:design:mutation:inst_gen}

A key aspect of the LLVM backend is to convert the wide range of LLVM IR types to the (usually small) set of types natively implemented by each target architecture.
Therefore, to exercise all the features of code generation, it is necessary to generate \ir{s} with as many data types as possible.
\sub{
    Many \ir{s} only operate on a restricted set of data types, and FuzzMutate only modelled scalar types, which is trivial and limited.
}
\add{
    Much of the compiler backend handles \ir{s} with vector types, but FuzzMutate modelled only scalar types.
}

\myComment{This address reviewer A(3): ``a discussion on the correctness of instruction modeling''.}
\sub{To overcome these limitations, we rewrite FuzzMutate's modelling to vector types.}
\add{
    
\begin{table*}[t]
    \centering
    \caption{
        \add{Extended instruction modeling for \ir{s}. Note that FuzzMutate only implements binary and bitwise operations with no vector support.}
    }
    \resizebox{\linewidth}{!}{
        \footnotesize
        \begin{tabular}{lrclll}
            \toprule
            Operation type                        & \multicolumn{1}{c}{Opcode}        &   & \multicolumn{3}{c}{Argument descriptions}                                                               \\
            \midrule
            \multirow{1}{*}{Unary operation}      & fneg                              & : & \multicolumn{3}{l}{anyFloatPointOrVectorFloatPoint}                                                     \\
            \cmidrule(lr){1-6}
            \multirow{2}{*}{Binary operations}    & add, sub, mul, (s$|$u)(div$|$rem) & : & anyIntOrVecInt                                      & sameAsFirst                       &               \\
                                                  & fadd, fsub, fmul, fdiv, frem      & : & anyFPOrVecFP                                        & sameAsFirst                       &               \\
            \cmidrule(lr){1-6}
            \multirow{1}{*}{Bitwise operations}   & shl, lshr, ashr, and, or, xor     & : & anyIntOrVecInt                                      & sameAsFirst                       &               \\
            \cmidrule(lr){1-6}
            \multirow{3}{*}{Vector operations}    & extractelement                    & : & anyVector                                           & anyInt                            &               \\
                                                  & insertelement                     & : & anyVector                                           & matchScalarOfFirst                & anyInt        \\
                                                  & shufflevector                     & : & anyVector                                           & matchLengthOfFirst                & VecOfConstI32 \\
            \cmidrule(lr){1-6}
            \multirow{2}{*}{Aggregate operations} & extractvalue                      & : & anyAggregateOrArray                                 & anyConstInt                       &               \\
                                                  & insertvalue                       & : & anyAggregateOrArray                                 & matchScalarOfFirst                & anyConstInt   \\
            \cmidrule(lr){1-6}
            \multirow{1}{*}{Memory operation}     & getelementptr                     & : & anySized                                            & pointerOfFirst                    & anyInt        \\
            \cmidrule(lr){1-6}
            \multirow{8}{*}{Casting operations}   & trunc                             & : & anyNonBoolIntOrVecInt                               & anyIntOrVecIntWithLowerPrecision  &               \\
                                                  & zext, sext                        & : & anyIntOrVecInt                                      & anyIntOrVecIntWithHigherPrecision &               \\
                                                  & fptrunc                           & : & anyNonHalfFPOrVecFP                                 & andFPOrVecFPWHigherPrecision      &               \\
                                                  & fptoui, fptosi                    & : & anyFPOrVecFP                                        & matchLengthOfFirstWithInt         &               \\
                                                  & uitofp, sitofp                    & : & anyIntOrVecInt                                      & matchLengthOfFirstWithFP          &               \\
                                                  & ptrtoint                          & : & anyPtrOrVecPtr                                      & matchLengthOfFirstWithInt         &               \\
                                                  & ptrtoint                          & : & anyIntOrVecInt                                      & matchLengthOfFirstWithPtr         &               \\
                                                  & bitcast                           & : & anyType                                             & anyTypeWithSameBitWidth           &               \\
            \cmidrule(lr){1-6}
            \multirow{3}{*}{Other operations}     & icmp                              & : & anyIntOrVecInt                                      & sameAsFirst                       &               \\
                                                  & fcmp                              & : & anyFPOrVecFP                                        & sameAsFirst                       &               \\
                                                  & select                            & : & anyBoolOrVecBool                                    & matchLengthOfFirst                & sameAsSecond  \\
            \bottomrule
        \end{tabular}
    }
    \vspace{-1.2\baselineskip}
    \label{table:inst_model}
\end{table*}

    To overcome FuzzMutate's limitations, we rewrite its modelling, as shown in~\autoref{table:inst_model}.
    We not only include vectors as allowed types but also model vector operations and casting operations.
}
These definitions are reflected in the code as declarations expressing both restrictions on the types of operands and constraints between the types of different operands.
For example, the \texttt{anyIntOrVecInt} constraint restricts the valid types for a particular operand to be any integer type or vector of integer type.
\sub{The \texttt{anyVector} constraint restricts the type of operand must be a vector of arbitrary length and element type.}
This allows us to model vector operations, such as \texttt{extractelement}, \texttt{insertelement}, and \texttt{shufﬂevector}, which were unsupported by FuzzMutate.


\sub{
    When generating a new instruction, we first randomly select an opcode and use the declarations to randomly select values that exist in the code with a compatible type.
    If no value exists with a compatible type, then the mutator will create a new operation with a compatible type.
    For numerical types, the new operation could generate a random constant, undef, or poison.
}

\sub{In addition, a small number of operations are not modeled declaratively.}
\add{In addition, }
\texttt{store} and \texttt{load} memory operations are structured differently enough from other operations that modeling them declaratively is unnecessary.
Some other instructions have constraints which are too complex \sub{to be simply handled in the declarative framework}, so we resort to custom generators.
For instance, instructions representing PHI nodes must be created with a number of operands equal to the number of predecessor blocks and must occur at the start of their basic block.
Similarly, \textit{call} instructions are handled manually, since we must select a function declaration and find values that exactly match the operand types of the declaration.

\add{
    When generating a new instruction, we first randomly select an opcode and use the declarations to randomly select values that exist in the code with a compatible type.
}
To ensure that values are defined before they are used, the mutator searches for values defined in the following locations: global variables, function arguments, values in dominators, and values defined by previous instructions in the same basic block.
If no value with a compatible type exists, then the mutator can \add{either generate a poison or } generate a load from a pointer if one exists.
\myComment{Since LLVM 17, LLVM has dropped typed pointers and started to use opaque pointers, thus the notion ``compatiable pointer type'' doesn't exist anymore.}
\sub{
    Lastly, if a value with a compatible pointer type exists, the mutator will fall back to either creating a new global variable, a new constant value, or a load from a stack memory location.
}

\myComment{This is a minor detail that doesn't seem important.}
\sub{
    In some cases, the mutator may create \ir{s} that define values which are never used.
    Since such dead code is likely to be removed by the compiler before instruction selection, the mutator will attempt to create a use for such values.
    One possibility is to store dead values to the stack or a global variable.
    Alternatively, if there are instructions after the definition, or the current block dominates other blocks, the mutator may select an instruction with a compatible operand to replace.
}

When generating instructions, the mutator may allocate new stack memories as placeholders.
To avoid undefined behaviors, the mutator will again attempt to replace loads from these placeholders with other values of a compatible type.
If no such value exists, then the mutator will store a value into the placeholder location.

We model no intrinsic functions, as they vary from architecture to architecture, potentially consuming much time with little outcome.
Instead, we rely on the feedback from matcher table coverage (\autoref{sec:design:coverage:feedback}), which shows the intrinsics that haven't been generated yet. The mutator will then randomly generate \textit{call} instructions to those intrinsics.

\add{
    \subsubsection{Instruction shuffling}
    \label{sec:design:mutation:shuffle}

    Changing instruction orders inside a basic block will change how the backend schedules instructions.
    When shuffling instructions, we must carefully handle instruction orders; otherwise, a use-before-definition may arise. We use topological sort to ensure that for each define-use relation, define precedes use after instruction shuffling.
}

\subsection{Matcher table feedback}
\label{sec:design:coverage}

\subsubsection{Matcher table instrumentation}
\label{sec:design:coverage:compression}

\add{
    LLVM uses \textit{patterns} to describe rewrite rules applied during instruction selection.
    Some simple patterns replace a single LLVM IR instruction with a single machine instruction.
    More complex patterns may replace multiple LLVM IR instructions or generate multiple machine instructions.
    Patterns may also apply in specific situations by including complex predicates.
    For example, a pattern may only apply when a particular operand is a constant, or a certain hardware feature is enabled.


    Most patterns are described declaratively in an LLVM-specific language called TableGen~\cite{TableGen}.
    To optimize the application of patterns, TableGen represents patterns in a state-machine and implements it as a large byte array known as the \textit{matcher table}.
    During compilation, the state machine determines the best pattern to apply to each \ir{}.
    \autoref{lst:codegen} is a C++ code snippet for evaluating the matcher table in \dagisel{}.
    \texttt{SDNode} is a data structure that represents an \ir{}.
    The while loop iteratively reads a command from the matcher table based on the current state, represented by the \texttt{idx} variable, evaluates the command, and selects the next state to be evaluated.
    For example, \texttt{Opc\_CheckOpcode} will check if the opcode of a given \texttt{SDNode} representing an instruction in the \dagisel{} graph matches a particular opcode.
    The \texttt{Result} will be used in future iterations, depending on the next entry in the matcher table.
    The compiler continues to evaluate the matcher table until it selects a single pattern or reaches a state where no pattern applies.

    \begin{lstlisting}[
    belowskip=-1.2\baselineskip,
    float, 
    language=C++, 
    caption={
        \dagisel{} in LLVM that consumes a matcher table to do instruction selection.
        \add{
          We also show AArch64's matcher table from index 25929 to 25936.
          Switch case \texttt{OPC\_MoveChild0} can be executed with different \texttt{Opc}, rendering branch coverage ineffective to track the behavior of this code.
          Therefore, we also track individual entries of the matcher table.
        }
    }, 
    label=lst:codegen,
    basicstyle=\ttfamily\scriptsize
] 
void SelectCodeCommon(SDNode *N, char *MatcherTable) {
 bool Result = true;
 unsigned Opc;
 while (true) {
  if (!Result) break;
  switch (MatcherTable[Idx++]){
    case OPC_CheckOpcode: {
     uint16_t Opc = MatcherTable[Idx++];
     Opc |= (unsigned short) MatcherTable[Idx++] << 8;
     Result = (Opc == N->getOpcode()); 
    }
    ...
    case OPC_MoveChild0: {
      unsigned ChildNo = Opc - OPC_MoveChild0;
      if (ChildNo >= N.getNumOperands())
        break;  // Match fails if out of range child #.
      N = N.getOperand(ChildNo);
      NodeStack.push_back(N);
      continue;
    }
  }
 }
}
void AArch64SelectionDAG::SelectCode(SDNode *N){
 #define TARGET_VAL(X) X & 255, unsigned(X) >> 8
 static const unsigned char MatcherTable[] = {
 ...
 /*25929*/OPC_CheckOpcode,TARGET_VAL(ISD::ADD),
 /*25932*/OPC_MoveChild0,
 /*25933*/OPC_CheckOpcode,TARGET_VAL(AArch64ISD::UMULL),
 /*25936*/OPC_MoveChild0,
 ...
 };
 SelectCodeCommon(N, MatcherTable,sizeof(MatcherTable));
}
\end{lstlisting}

    Since the program in \autoref{lst:codegen} evaluates all the patterns using the same set of conditional branches in the switch statement, its control flow coverage does not reflect what patterns have been exercised.
}
\sub{
    \Mir{} generation does not correspond to compiler control flow.
    The same control flow can be used to generate different instructions due to the matcher table design, as shown in \autoref{lst:codegen}.
    Consequently, many patterns may not be generated when edge coverage is high.
}
To overcome this, we track the usage of the matcher table directly.

\myComment{Addressing Reviewer A(4): ``... the concept of utilizing matcher table ... lacks clarity''}
\sub{
    Similar to edge coverage, we allocate a table when the compiler starts in order to track the coverage of the matcher table.
    Every time an entry in the matcher table is accessed, we will record that access in our table as well.
}

\add{
    Similar to how AFL tracks branch coverage, we allocate a table, \emph{matcher table coverage table}, for tracking the coverage of the matcher table.
    Each entry in this table corresponds to an entry in the matcher table and records if the latter has been accessed. The instrumented compiler dumps matcher table coverage after every execution.
    If either the branch coverage table and or the matcher table coverage table shows new coverage, then the fuzzer considers the input as new.
}

\begin{table}[t]
    \centering
    \caption{
        The number of entries in the matcher tables used by SelectionDAG in mature architectures (LLVM commit \LLVMVersion).
        \add{To track the coverage of the matcher table, we use one bit to track each entry in the matcher table.}
    }
    \resizebox{\linewidth}{!}{
        \begin{tabular}{lS[table-format=7]||lS[table-format=7]}
            \toprule
            Arch    & {\# of entries}  & Arch    & {\# of entries}  \\
            \midrule
            AArch64 & 489789       &  PowerPC & 190304        \\
            AMDGPU  & 493556       &  RISC-V   & 2191899      \\
            ARM     & 201172       &  SystemZ & 53271         \\
            Hexagon & 178277       &  VE      & 71577         \\
            Mips    & 54044        &  WASM    & 25991         \\
            NVPTX   & 186134       &  X86     & 680916        \\
            \bottomrule
        \end{tabular}
    }
    \vspace{-1\baselineskip}
    \label{table:matcher_table_size}
\end{table}

Tracking matcher table coverage incurs memory overhead, which may reduce fuzzing throughput~\cite{241986}.
The second and fourth column of \autoref{table:matcher_table_size} show the size of the matcher table in different mature architectures.
The matcher tables for mainstream architectures, such as X86 and AArch64, have several hundred thousand entries, whereas RISC-V has about two million entries.
\myComment{Addressing Reviewer C: ``Does IRFuzzer take into consideration the order of elements that have been accessed in the matcher tables?''}
\sub{
    However, unlike control flow where the edge's execution count represents different program semantics, a matcher table entry being accessed repeatedly only means the same pattern is triggered multiple times.
    Therefore, we only track whether an entry is accessed or not, i.e., we use a boolean to track each entry.
}
\add{
    Since the entries in the matcher table represent different features, to determine which features have been covered, we can individually track whether each entry has been accessed because the order of access is irrelavent. To reduce memory footprint, we use one byte to track eight entries in the matcher table. For example, the largest matcher table, of the RISC-V architecture, has \num{2191899} entries. Tracking its coverage takes $\lceil{2191899 / 8}\rceil$ bytes, or 274 kB.
}
\sub{
    This observation gives us a chance to optimize our instrumentation and thus reduce the memory overhead.
    During instrumentation, we pack eight booleans into a byte to save space.
    If the table size is not a multiple of eight, we pad extra booleans.
}

\sub{
    During fuzzing, to access an entry in the matcher table, the fuzzer can calculate the offset of the entry's corresponding boolean using its index.
    After execution, the instrumented compiler will report a matcher table coverage back to the fuzzer.
    The fuzzer will use edge coverage \textit{and} matcher table coverage together.
    If either table shows new coverage, we will consider the input as new.
}

\sub{
    \subsubsection{IR mutation feedback}
    \label{sec:design:coverage:feedback}
    While the matcher table can help filter out seeds with no new behavior, it also contains knowledge whether an instruction or intrinsic is generated or not.
}
\add{
    \subsubsection{IR mutation feedback}
    \label{sec:design:coverage:feedback}
    The mutator needs to know which patterns in the matcher table have been executed so that it can generate more diverse inputs. However, when LLVM prepares the matcher table, it hides which pattern each entry in the matcher table represents.
}

\myComment{Addressing Reviewer A(4): ``crucial details regarding the structure and content of this table are absent''}
\sub{
    We first modify TableGen to dump a look-up table specifying the correspondence of matcher table entries and machine instruction patterns.
    The pattern reveals the condition on a specific instruction or intrinsic being generated.
}
\add{
    To recover this information, we generate a look-up table to map each matcher table entry to its corresponding machine instruction pattern.
    Compiler developers program different patterns into TableGen, and the compiler translates those patterns into the matcher table.
    We modify TableGen to reverse that process to create the look-up table.
}

Prior to fuzzing, we create this look-up table for each architecture.
During fuzzing, we use the matcher table coverage table and the look-up table to determine which patterns haven't been generated. 
Finally, we send this report to the mutator to encourage it to generate those patterns, which is done every ten minutes to avoid excessive runtime overhead.

\section{Implementation}
\label{sec:impl}

\myComment{Modify this to answer Reviewer A(5): ``he paper hints at an expansion of the mutation component ... a clearer differentiation between the two ... would enhance understanding.''}
\add{
    Our implementation is based on prior work \fuzzmutate{}\cite{FuzzMutate} and AFL++~\cite{AFLplusplus}.
}
\add{
    Compared with FuzzMutate, we added the following new mutation strategies, which have been incorporated into the upstream LLVM's repository:
    \begin{itemize}
        \item A new function template generator with the ability to modify function attributes.
        \item A new control flow graph mutation strategy, \sCFG insertion strategy, which modifies the control flow while preserving domination relations.
        \item Extended modelling of IR instructions, including PHI nodes, memory operations, vector operations, and support for non-scalar types.
    \end{itemize}
    Compared with AFL++, we measure the coverage of the matcher table, which
    not only helps determine if a new input is interesting but also guides mutation. This feedback allows our mutator to generate architecture specific intrinsics without any prior knowledge of the architecture.
}

\section{Evaluation}
\label{sec:eval}

We evaluated \irfuzzer by fuzzing LLVM with different settings and tools to answer the following research questions:

\begin{itemize}
    \item \textbf{RQ1}: How does \irfuzzer{} compare with \sota backend fuzzers?
    \item \textbf{RQ2}: How does \irfuzzer{} compare with end-to-end fuzzers like CSmith and GrayC?
    \item \textbf{RQ3}: Do mutator and matcher table feedback individually contribute to \irfuzzer{}?
    \item \textbf{RQ4}: Can \irfuzzer{} find new bugs in LLVM?
\end{itemize}

The upstream LLVM repository (commit \LLVMVersion{}) currently supports 21 architectures.
We only tested on mature architectures that had a matcher table size larger than \num{25000}, as shown in ~\autoref{table:matcher_table_size}.
In addition, each architecture may provide different features that can be enabled on different hardware.
For simplicity, we selected the backends of some popular microchips, which had a predefined set of features.
These backends were widely used from user product to server applications, justifying the variety of our choice.
All the architectures that we tested were under active development.
As a result, we selected \numCPU{} \tCPU{s}\footnote{``\TCPU{}'' was used in LLVM to label a backend corresponding to a microchip. It can also refer to GPU, DSP or virtual targets like WebAssembly.} across \numArch{} architectures.

We used two baseline fuzzers: (1) AFL++ with no modification, and (2) AFL++ whose mutation module was replaced with \fuzzmutate{}, referred to as \fuzzmutate{} thereafter.
All fuzzers used AFL++'s default scheduling.
For fairness, we collected the seeds generated by each fuzzer and measured their branch coverage and matcher table coverage.
AFL++ reported branch coverage using classical instrumentation and a default 64 kB table.
\myComment{Reviewer A(2): `` the methodology behind ... percentages, including the denominators, remains unclear''. TODO: If Henry's expr can come through, we can report line coverage.}

We prepared two versions of \irfuzzer{}:
\begin{itemize}
\item \irfuzzer has all the mechanisms described in \autoref{sec:design}.
\item \irfuzzer{}$_\text{bare}$ excludes the feedback mechanism described in \autoref{sec:design:coverage}. 
Its performance reveals the contribution of our mutator when compared with \fuzzmutate{}, and of the feedback mechanism when compared with \irfuzzer{}.
\end{itemize}

Each fuzzer process ran exclusively on a single processor core on an x86\_64 server.
Each fuzzing process ran for one day to allow adequate exploration~\cite{10.1145/3243734.3243804}.
We repeated each experiment five times to average the results to reduce random effects.
To demonstrate \irfuzzer{}'s ability to mutate IR modules and to provide a fair comparison with AFL++, we initialized each fuzzer process with 92 seeds.
We randomly selected the seeds from LLVM's unit tests. Each seed was smaller than 256 bytes to increase the throughput. We anonymously published the seeds in the artifact~\cite{IrfuzzerArtifact}.

\subsection{Baseline comparison}
\label{sec:eval:baseline}

We compared our mutation strategy with two baseline implementations: AFL++ and the upstream LLVM implementation of \fuzzmutate{}.
AFL++ lacks an LLVM IR-aware mutator, whereas \fuzzmutate{} has a limited LLVM IR-aware mutator.

\begin{table*}[t]
    \centering
    \caption{
        Branch table coverage and matcher table coverage on \numCPU{} \tCPU{s} across \numArch{} targets in \dagisel{}.
        Statistics are the arithmetic mean over five trials.
        Bold entries are the best among baseline fuzzers.
        FM means AFL++ coupled with FuzzMutate, IRF means \irfuzzer{}, IRF$_{\text{bare}}$ means \irfuzzer{} without matcher table feedback.
    }
    \sisetup{table-format=2.2}
    \resizebox{0.95\linewidth}{!}{
        \begin{tabular}{llrrrrrrrrrr}
            \toprule
            \multirow{2}{*}{Arch}               &
            \multirow{2}{*}{\TCPU{}}            &
            \multicolumn{5}{c}{Branch coverage} &
            \multicolumn{5}{c}{Matcher table coverage}                                                                                                                                                                                                                                                        \\
            \cmidrule(lr){3-7} \cmidrule(lr){8-12}
                                                &                & Seeds             & AFL++             & FM                & IRF$_{\text{bare}}$        & IRF                        & Seeds             & AFL++             & FM                & IRF$_{\text{bare}}$        & IRF                         \\
            \midrule
            \multirow{7}{*}{AArch64}            & apple-a16      & 59.8\si{\percent} & 87.1\si{\percent} & 82.9\si{\percent} & 95.2\si{\percent}          & \bfseries96.9\si{\percent} & 0.7\si{\percent}  & 1.6\si{\percent}  & 2.6\si{\percent}  & 7.5\si{\percent}           & \bfseries8.9\si{\percent}   \\
                                                & apple-m2       & 59.8\si{\percent} & 86.9\si{\percent} & 83.3\si{\percent} & 94.9\si{\percent}          & \bfseries97.0\si{\percent} & 0.7\si{\percent}  & 1.6\si{\percent}  & 2.6\si{\percent}  & 7.6\si{\percent}           & \bfseries9.2\si{\percent}   \\
                                                & cortex-a715    & 60.0\si{\percent} & 87.7\si{\percent} & 83.2\si{\percent} & 94.9\si{\percent}          & \bfseries96.9\si{\percent} & 0.7\si{\percent}  & 1.7\si{\percent}  & 2.6\si{\percent}  & 7.4\si{\percent}           & \bfseries10.9\si{\percent}  \\
                                                & cortex-r82     & 60.1\si{\percent} & 87.0\si{\percent} & 82.9\si{\percent} & 95.2\si{\percent}          & \bfseries96.7\si{\percent} & 0.7\si{\percent}  & 1.6\si{\percent}  & 2.6\si{\percent}  & 7.3\si{\percent}           & \bfseries8.8\si{\percent}   \\
                                                & cortex-x3      & 60.0\si{\percent} & 93.3\si{\percent} & 85.2\si{\percent} & 96.6\si{\percent}          & \bfseries96.8\si{\percent} & 0.7\si{\percent}  & 7.1\si{\percent}  & 2.7\si{\percent}  & 7.9\si{\percent}           & \bfseries10.5\si{\percent}  \\
                                                & exynos-m5      & 60.3\si{\percent} & 87.4\si{\percent} & 83.2\si{\percent} & \bfseries96.5\si{\percent} & 96.2\si{\percent}          & 0.7\si{\percent}  & 1.7\si{\percent}  & 2.6\si{\percent}  & 7.9\si{\percent}           & \bfseries8.5\si{\percent}   \\
                                                & tsv110         & 60.0\si{\percent} & 87.3\si{\percent} & 82.9\si{\percent} & \bfseries95.9\si{\percent} & 95.7\si{\percent}          & 0.7\si{\percent}  & 1.6\si{\percent}  & 2.6\si{\percent}  & 7.7\si{\percent}           & \bfseries8.2\si{\percent}   \\
            \specialrule{.4pt}{1pt}{1pt}
            \multirow{2}{*}{AMDGPU}             & gfx1036        & 70.8\si{\percent} & 90.0\si{\percent} & 89.1\si{\percent} & 96.2\si{\percent}          & \bfseries97.0\si{\percent} & 0.9\si{\percent}  & 2.1\si{\percent}  & 2.7\si{\percent}  & 4.3\si{\percent}           & \bfseries 5.1\si{\percent}  \\
                                                & gfx1100        & 71.2\si{\percent} & 89.7\si{\percent} & 89.9\si{\percent} & 96.6\si{\percent}          & \bfseries96.8\si{\percent} & 1.0\si{\percent}  & 2.1\si{\percent}  & 2.9\si{\percent}  & 4.4\si{\percent}           & \bfseries4.9\si{\percent}   \\
            \specialrule{.4pt}{1pt}{1pt}
            \multirow{1}{*}{ARM}                & generic        & 55.5\si{\percent} & 87.9\si{\percent} & 82.5\si{\percent} & 88.6\si{\percent}          & \bfseries91.6\si{\percent} & 1.7\si{\percent}  & 4.3\si{\percent}  & 4.3\si{\percent}  & 4.3\si{\percent}           & \bfseries5.4\si{\percent}   \\

            \specialrule{.4pt}{1pt}{1pt}
            \multirow{2}{*}{Hexagon}            & hexagonv71t    & 64.8\si{\percent} & 88.0\si{\percent} & 86.0\si{\percent} & 93.2\si{\percent}          & \bfseries94.8\si{\percent} & 1.7\si{\percent}  & 6.6\si{\percent}  & 17.0\si{\percent} & 21.6\si{\percent}          & \bfseries 33.2\si{\percent} \\
                                                & hexagonv73     & 64.9\si{\percent} & 89.5\si{\percent} & 85.7\si{\percent} & 93.0\si{\percent}          & \bfseries94.7\si{\percent} & 1.7\si{\percent}  & 7.3\si{\percent}  & 17.4\si{\percent} & 20.7\si{\percent}          & \bfseries32.5\si{\percent}  \\
            \specialrule{.4pt}{1pt}{1pt}
            \multirow{1}{*}{Mips}               & mips64r6       & 52.5\si{\percent} & 81.0\si{\percent} & 72.7\si{\percent} & \bfseries87.0\si{\percent} & 84.8\si{\percent}          & 3.8\si{\percent}  & 10.0\si{\percent} & 15.3\si{\percent} & \bfseries18.4\si{\percent} & 18.3\si{\percent}           \\
            \specialrule{.4pt}{1pt}{1pt}
            \multirow{1}{*}{NVPTX}              & sm\_90         & 46.6\si{\percent} & 77.5\si{\percent} & 77.5\si{\percent} & 90.6\si{\percent}          & \bfseries91.3\si{\percent} & 1.7\si{\percent}  & 3.1\si{\percent}  & 4.7\si{\percent}  & 6.3\si{\percent}           & \bfseries26.9\si{\percent}  \\
            \specialrule{.4pt}{1pt}{1pt}
            \multirow{1}{*}{PowerPC}            & pwr9           & 60.3\si{\percent} & 87.3\si{\percent} & 86.9\si{\percent} & 95.6\si{\percent}          & \bfseries95.9\si{\percent} & 1.2\si{\percent}  & 3.6\si{\percent}  & 7.1\si{\percent}  & 19.0\si{\percent}          & \bfseries23.6\si{\percent}  \\
            \specialrule{.4pt}{1pt}{1pt}
            \multirow{3}{*}{RISC-V}              & rocket-rv64    & 53.7\si{\percent} & 83.0\si{\percent} & 76.6\si{\percent} & 87.1\si{\percent}          & \bfseries88.3\si{\percent} & 0.12\si{\percent} & 0.20\si{\percent} & 0.22\si{\percent} & 0.23\si{\percent}          & \bfseries0.23\si{\percent}  \\
                                                & sifive-u74     & 54.5\si{\percent} & 83.1\si{\percent} & 75.9\si{\percent} & \bfseries88.3\si{\percent} & 88.2\si{\percent}          & 0.14\si{\percent} & 0.24\si{\percent} & 0.29\si{\percent} & 0.31\si{\percent}          & \bfseries0.32\si{\percent}  \\
                                                & sifive-x280    & 55.0\si{\percent} & 84.1\si{\percent} & 75.7\si{\percent} & 90.7\si{\percent}          & \bfseries92.0\si{\percent} & 0.14\si{\percent} & 0.27\si{\percent} & 0.31\si{\percent} & 3.42\si{\percent}         & \bfseries3.70\si{\percent} \\
            \specialrule{.4pt}{1pt}{1pt}
            \multirow{2}{*}{SystemZ}            & z15            & 55.3\si{\percent} & 84.0\si{\percent} & 81.5\si{\percent} & 93.7\si{\percent}          & 93.8\si{\percent}          & 5.2\si{\percent}  & 13.7\si{\percent} & 27.1\si{\percent} & 43.9\si{\percent}          & \bfseries50.6\si{\percent}  \\
                                                & z16            & 55.3\si{\percent} & 83.7\si{\percent} & 81.8\si{\percent} & 93.3\si{\percent}          & \bfseries93.7\si{\percent} & 5.2\si{\percent}  & 14.1\si{\percent} & 26.5\si{\percent} & 43.7\si{\percent}          & \bfseries50.2\si{\percent}  \\
            \specialrule{.4pt}{1pt}{1pt}
            \multirow{1}{*}{VE}                 & generic        & 49.0\si{\percent} & 80.4\si{\percent} & 70.2\si{\percent} & \bfseries89.6\si{\percent} & 89.0\si{\percent}          & 3.5\si{\percent}  & 8.1\si{\percent}  & 11.4\si{\percent} & 13.0\si{\percent}          & \bfseries14.1\si{\percent}  \\
            \specialrule{.4pt}{1pt}{1pt}
            \multirow{2}{*}{WASM}               & bleeding-edge  & 46.8\si{\percent} & 84.7\si{\percent} & 70.5\si{\percent} & 88.8\si{\percent}          & \bfseries90.0\si{\percent} & 4.1\si{\percent}  & 36.9\si{\percent} & 10.9\si{\percent} & 40.2\si{\percent}          & \bfseries41.5\si{\percent}  \\
                                                & generic        & 46.6\si{\percent} & 80.2\si{\percent} & 69.7\si{\percent} & 87.4\si{\percent}          & \bfseries88.4\si{\percent} & 4.1\si{\percent}  & 11.8\si{\percent} & 10.6\si{\percent} & 12.0\si{\percent}          & \bfseries12.4\si{\percent}  \\
            \specialrule{.4pt}{1pt}{1pt}
            \multirow{6}{*}{X86}                & alderlake      & 61.2\si{\percent} & 88.0\si{\percent} & 84.6\si{\percent} & 96.3\si{\percent}          & \bfseries97.2\si{\percent} & 0.7\si{\percent}  & 1.8\si{\percent}  & 3.1\si{\percent}  & 7.1\si{\percent}           & \bfseries9.3\si{\percent}   \\
                                                & emeraldrapids  & 60.5\si{\percent} & 93.4\si{\percent} & 84.4\si{\percent} & 96.2\si{\percent}          & \bfseries97.5\si{\percent} & 0.6\si{\percent}  & 12.5\si{\percent} & 3.2\si{\percent}  & 14.8\si{\percent}          & \bfseries18.9\si{\percent}  \\
                                                & raptorlake     & 61.2\si{\percent} & 93.5\si{\percent} & 85.8\si{\percent} & 96.8\si{\percent}          & \bfseries97.2\si{\percent} & 0.7\si{\percent}  & 6.2\si{\percent}  & 3.3\si{\percent}  & 7.4\si{\percent}           & \bfseries9.4\si{\percent}   \\
                                                & sapphirerapids & 60.5\si{\percent} & 88.4\si{\percent} & 85.4\si{\percent} & 96.7\si{\percent}          & \bfseries97.4\si{\percent} & 0.6\si{\percent}  & 1.8\si{\percent}  & 3.3\si{\percent}  & 15.3\si{\percent}          & \bfseries19.1\si{\percent}  \\
                                                & znver3         & 61.8\si{\percent} & 86.6\si{\percent} & 84.0\si{\percent} & 96.5\si{\percent}          & \bfseries97.4\si{\percent} & 0.7\si{\percent}  & 1.6\si{\percent}  & 3.0\si{\percent}  & 7.3\si{\percent}           & \bfseries9.3\si{\percent}   \\
                                                & znver4         & 61.0\si{\percent} & 87.6\si{\percent} & 84.0\si{\percent} & 96.3\si{\percent}          & \bfseries97.5\si{\percent} & 0.7\si{\percent}  & 1.8\si{\percent}  & 3.2\si{\percent}  & 14.4\si{\percent}          & \bfseries17.7\si{\percent}  \\
            \bottomrule
        \end{tabular}
    }
    \vspace{-1\baselineskip}
    \label{table:baseline}
\end{table*}

\autoref{table:baseline} shows the branch and matcher table coverages, which we calculated by dividing the number of non-empty entries in the coverage table by the size of the table. The \emph{seeds} columns show the coverage brought by the initial seeds.
\sub{
On each line, the \textbf{bold} numbers are the best statistical significance ($p < 0.05$) when compared with other baseline fuzzers using Mann Whitney U Test.
}
\add{
On each \tCPU{}, Target CPU, \irfuzzer{} and \irfuzzer{}$_\text{bare}$ achieved more coverage than the baseline fuzzers, and the difference is statistically significant ($p < 0.05$).
}

\sub{
    Overall, we see AFL++ performed poorly for the purpose of testing LLVM compiler backends.
    In most backends, it cannot increase much matcher table coverage due to its lack of support for structured input; \irfuzzer{} covered more branches than AFL++ on 28 \tCPU{s}.
    The output generated by AFL++ did not provide significant coverage of instruction selection patterns, as measured by the low matcher table coverage.
}
\add{
    \irfuzzer{} achieved the highest branch coverage on all the \tCPU{s}. It may seem counterintuitive that AFL++ has higher branch coverage than \fuzzmutate{} on most \tCPU{s}. Our investigation revealed that AFL++'s high branch coverage mostly comes from error handling code since it can hardly generate valid input. This is further demonstrated by AFL++'s low matcher table coverage, which indicates that most executions did not reach the instruction selection stage before the compiler terminated.
}

\sub{
    Both \fuzzmutate{} and \irfuzzer{} reached high code coverages.
    \fuzzmutate{} reached more than 75\% except for mips64r6, generic VE, and generic WebAssembly.
    On the other hand, \irfuzzer{} performed better, achieving over 80\% branch coverage for all \tCPU{s}.
}

It is insufficient to compare only branch coverage~\cite{8418631}.
More significantly, \irfuzzer{} achieved the best matcher table coverage on all CPUs, indicating significantly better coverage of instruction selection patterns.
\sub{
    We observe that in generic WebAssembly, \irfuzzer{} shows no significance compared with other fuzzers.
    After investigation, we find that generic WebAssembly disabled many features, limiting the maximum reachable matcher table to 11.8\%.
    This does not show that \irfuzzer{} is less effective on WebAssembly, as we can see that \irfuzzer{} still rank number one in the bleeding-edge version.
}

\add{
    Comparison between \fuzzmutate{} and AFL++ also cast insights on which fuzzer is better to fuzz the backend compiler.
    \fuzzmutate{} can generate valid input to reach deeply nested code more easily, as demonstrated by its higher matcher table coverage in \autoref{table:baseline} compared with AFL++.
    On the other hand, AFL++'s high branch coverage and low matcher table coverage show that most inputs didn't reach the instruction selection stage before the compiler terminated.
    Therefore, AFL++ is useful mainly for testing error handling and the frontend.
}

In summary, \irfuzzer{} achieved higher branch coverage and matcher table coverage on \sub{28 out of \numCPU{}} \add{all} \tCPU{s} \add{compared with AFL++ and \fuzzmutate{}}.
To answer \textbf{RQ1}, \irfuzzer{} is better in coverage when fuzzing LLVM code generation compared with \sota fuzzers.

\subsection{Comparison with end-to-end fuzzers}
\label{sec:eval:csmith}

To better understand the benefits of targeted fuzzing over end-to-end fuzzing, we evaluated CSmith~\cite{CSmith} and GrayC~\cite{GreyC}.
Unlike \irfuzzer{}, end-to-end fuzzers generate C code, which must be processed by the compiler frontend and middle-end before reaching the backend.
As a result, they exercise the entire compilation pipeline, rather than focusing on just the backend.
Note that although CSmith generates random, syntactically correct C code, it does not implement any instrumentation and lacks feedback to guide the generation process.
While GrayC relies on branch coverage feedback, it does not have feedback that is customized for the backends of the compilers.
\sub{
    Besides, to test end-to-end fuzzers, we have to cross compile C to different architectures.
    Cross compilation is difficult, as the testers have to set up the proper tool chain for it.
}
\add{Besides, to test end-to-end fuzzers, we had to cross compile C to different architectures, which was difficult and time-comsuming.}
Therefore, we tested on three most widely used architectures using generic backend.

CSmith generates C files with no initial seed.
To make the comparison fair, we also ran \irfuzzer{} with \textbf{no} initial seed, since \irfuzzer{} is capable of generating LLVM IR from scratch.
GrayC relies on deprecated APIs in LLVM 12 and cannot instrument the latest LLVM. So we download the artifact provided by GrayC~\cite{GrayCArtifact}.
The artifact consists of \num{715147} C programs across ten trials.
We ran CSmith for 24 hours and repeated it eight times, generating a total of \num{506971} C programs.

We cross-compiled these C programs to different architectures.
After compilation, we measured the resulting branch and matcher table coverage in the compiler backend, using the same instrumentation as \irfuzzer{}.
\sub{
    \texttt{llc} default optimization to \texttt{O2}, therefore we only test \texttt{O2} and \texttt{O3}, as \texttt{O0} and \texttt{O1} are often subsets of \texttt{O2}.
}
\add{
    We only tested on \texttt{O2} and \texttt{O3}, as \texttt{O0} and \texttt{O1} are often subsets of \texttt{O2}.
}
The results are shown in \autoref{table:csmith}.

\begin{table}[t]
    \centering
    \caption{
        Average branch table coverage and matcher table coverage of CSmith (CS), GrayC, and \irfuzzer{} (IRF).
        \texttt{O2} and \texttt{O3} are different optimization levels.
        Bold entries are the winners.
    }
    \sisetup{table-format=2.1}
    \begin{tabular}{llSSSSSS}
        \toprule
                                                  &
        \multirow{2}{*}{Arch}                     &
        \multicolumn{3}{c}{Branch table coverage} &
        \multicolumn{3}{c}{Matcher table coverage}                                                                                    \\
        \cmidrule(lr){3-5}\cmidrule(lr){6-8}
                                                  &         & {CS}   & {GrayC} & {IRF}           & {CS}  & {GrayC} & {IRF}            \\
        \midrule
        \multirow{3}{*}{\texttt{O2}}              & AArch64 & 94.8\% & 96.1\%  & \bfseries96.7\% & 5.2\% & 6.9\%   & \bfseries  8.9\% \\
                                                  & ARM     & 90.7\% & 92.3\%  & \bfseries92.5\% & 4.5\% & 4.5\%   & \bfseries  5.4\% \\
                                                  & X86     & 94.8\% & 96.1\%  & \bfseries96.9\% & 3.5\% & 4.2\%   & \bfseries  5.9\% \\
        \midrule
        \multirow{3}{*}{\texttt{O3}}              & AArch64 & 95.3\% & 96.2\%  & \bfseries96.9\% & 5.4\% & 6.9\%   & \bfseries 8.9\%  \\
                                                  & ARM     & 91.1\% & 92.5\%  & \bfseries92.5\% & 4.5\% & 4.5\%   & \bfseries 5.4\%  \\
                                                  & X86     & 94.9\% & 96.2\%  & \bfseries96.8\% & 3.5\% & 4.2\%   & \bfseries 5.9\%  \\
        \bottomrule
    \end{tabular}
    \vspace{-\baselineskip}
    \label{table:csmith}
\end{table}

\irfuzzer{} achieved \sub{higher}\add{the highest} matcher table \add{and branch } coverage on all the architectures and all the optimizations.
Even with branch coverage feedback, GrayC was unable to generate C inputs with more matcher table coverage, which demonstrating the need for specialized backend fuzzing.
\sub{
    We looked into the code generated by end-to-end fuzzers and found the reason for the coverage differences.
    The inferiority of matcher table coverage of end-to-end fuzzers was largely related to a lack of coverage of vector data types.
}
\add{
    We looked into the code generated by end-to-end fuzzers and found that  their low matcher table coverage was mainly because they could not handle vector data types.
}
Vector instructions are generated only when the front-end and middle-end decide that a vector instruction will speed up a particular piece of code, which is uncommon in random C programs generated by end-to-end fuzzers.
In comparison, since \irfuzzer{} operates directly on \ir{s}, it can generate vector operations easily.

\sub{
    On the other hand, CSmith and GrayC achieved higher branch coverage. After investigation, we find that the size of generated files contributed to the difference. \irfuzzer{} generates bit code less than 10 KB to achieve higher throughput. Seeds generated by CSmith average to more than 40 KB  after we compiled them to bit code.
    This means that many backend edges are executed more times, which would increase branch coverage in AFL's design.
    This does not show \irfuzzer{}'s inferiority. \irfuzzer{} can reach fair coverage in much smaller inputs, showing the efficiency of specialized backend fuzzing.
}

To answers \textbf{RQ2}: \irfuzzer{} achieved higher matcher table coverage than \sota end-to-end fuzzers.
This shows that compiler backend testing should not solely rely on end-to-end fuzzing, and that specialized fuzzing can improve matcher table coverage significantly.

\subsection{Individual contributions}
\label{sec:eval:matcher}

To evaluate how each component of \irfuzzer{} helps, we stripped all the feedbacks in \irfuzzer{} to get \irfuzzer{}$_{\text{bare}}$.

\autoref{table:baseline} shows that \irfuzzer{}$_{\text{bare}}$ \emph{always} reached higher branch coverage and matcher table coverage than \fuzzmutate, indicating that our mutator was able to generate more diverse inputs.
Although \fuzzmutate{} is also a structured mutator, it lacks many advanced features that we designed in \autoref{sec:design:mutation}.
The sifive-x280 CPU best demonstrates this improvement, where \irfuzzer{}$_{\text{bare}}$ covered 3.42\si{\percent} of the matcher table while \fuzzmutate{} covered only 0.31\si{\percent}.

The last two columns of \autoref{table:baseline} show that \irfuzzer{} is able to cover more matcher table in 28 out of \numCPU \tCPU{s} compared with \irfuzzer{}$_{\text{bare}}$.
\add{
    This demonstrates that our matcher table feedback can help the mutator during fuzzing trials.
    This effect can be best observed on NVPTX, where \irfuzzer{}$_{\text{bare}}$ covered only 6.3\si{\percent} of the matcher table while \irfuzzer{} covered 26.9\si{\percent}.
}
\sub{
    \irfuzzer{} didn't show superiority on three \tCPU{s} (rocket-rv64, sifive-x280, generic WebAssembly) mainly because both fuzzers reached coverage ceilings allowed by that \tCPU{}.
}

\sub{We find that sometimes \irfuzzer{} has less branch coverage.}
\add{In rare cases \irfuzzer{} has lower branch coverage than \irfuzzer{}$_\text{bare}$.} This is because the feedback mechanism incurs a tradeoff.
Calculating matcher table coverage and sending it to the mutator reduce the throughput, which lowers branch coverage.
On the other hand, this feedback is valuable for generating more diverse inputs, which contributes to higher matcher table coverage.
\sub{
    For example, in NVPTX, \irfuzzer{} achieved 26.5\% matcher table thanks to our feedback when \irfuzzer{}$_{\text{bare}}$ only covered 6.2\%.
    Since our end goal is to test the code generation part of the backend, we believe this tradeoff is acceptable.
}
\add{
    Among all the \numCPU \tCPU{s}, \irfuzzer had lower branch coverage than \irfuzzer{}$_{\text{bare}}$ on only 5 \tCPU{s}, so we believe that the tradeoff is acceptable and justified.
    Besides, both \irfuzzer and \irfuzzer{}$_{\text{bare}}$ outperformed baseline fuzzers on all the \tCPU{s}.
}
We can answer \textbf{RQ3} confidently that both the mutator and feedback mechanism contributed to improved matcher table coverage.

\subsection{Bug categories and analysis}
\label{sec:eval:types}

We collected all the crashes found in ~\autoref{sec:eval:baseline} and ~\autoref{sec:eval:csmith}.
We also fuzzed other architectures with no features to extend our scope.
Since \gisel also uses matcher table design, we can apply \irfuzzer{} on it with little modification.
This demonstrates that our approach can be generalized to other frameworks with little effort.

In the process, we found hundreds of crashes in the LLVM compiler.
Even though these crashes all have unique stack traces, they do not necessarily indicate different bugs because some crashes have different paths but the same root cause.
Therefore, we manually analyzed all of them and reported the ones that we believe are bugs.
In this section, we only report the bugs that have been \textit{confirmed}.
In total, \irfuzzer{} found \numTotalBugs{} confirmed bugs.
We manually verified that these bugs are found only by \irfuzzer{} and published the details anonymously~\cite{IrfuzzerArtifact}.


These bugs are distributed in different places in the LLVM codebase.
\autoref{fig:bugs:locations} shows the distribution of these bugs across LLVM.
CodeGen is the library shared between architectures, meaning that a bug in CodeGen may affect all architectures.

\myComment{We divide into three subsections to address meta review 1 and 3}
\add{
    \subsubsection{Bugs found by baseline fuzzers}
    \label{sec:eval:types:others}
    Our evaluation of the baseline fuzzer --- AFL++, \fuzzmutate{}, CSmith, and GrayC --- shows that none of them found any backend bugs. All the \numTotalBugs{} confirmed bugs were found exclusively by \irfuzzer{}.
    AFL++ found many crashes in the \texttt{llc} lexer and module verifier. However, all of them were caused by a malformed input and are not considered bugs.
    \fuzzmutate{} did not find any crashes because its mutator is very limited and only covers common use cases of the compiler backend.
}

\subsubsection{Distribution of bugs}
We categorize these bugs into six categories: hang, memory errors, assertion failures, logic errors, missing patterns, and other bugs.
Hang, memory errors and assertion failures are the most severe because they stall compilation.
A missing pattern bug occurs when a certain \mir{} is permitted by the hardware specification but no matching instruction selection pattern exists.
Logic errors and missing patterns do not stall compilation but may generate ineffective or even wrong \mir{s}.
\autoref{fig:bugs:catogories} shows the number of bugs in each category. 
Assertion bugs are the most common. They arise from the developers' false assumption that some properties hold during compilation, which our fuzzer disapproved.

\pgfplotstableread{
    Label fixed confirmed
    Assertion 18 4
    Pattern 16 6
    Logic 10 5
    Memory 7 3
    Hang 3 3
    Other 3 1
}\bugtypes

\pgfplotstableread{
    Label fixed confirmed
    X86 10 6
    AArch64 11 2
    CodeGen 9 2
    AMDGPU 5 3
    NVPTX 2 6
    Hexagon 5 2
    WASM 7 0
    RISC-V 3 0
    ARM 1 0
    BPF 1 0
    VE 1 0
    XCore 1 0
    PowerPC 1 0
}\buglocations

\begin{figure}[t]
    \centering
    \begin{subfigure}[t]{\linewidth}
        \centering
        \resizebox{\linewidth}{!}{
            \begin{tikzpicture}
                \begin{axis}[
                        width=\linewidth,
                        height=9\baselineskip,
                        ybar stacked,
                        ymin=0,
                        ymax=18,
                        xtick=data,
                        legend style={cells={anchor=west}, legend pos=north east, font=\scriptsize},
                        reverse legend=true, 
                        xticklabels from table={\buglocations}{Label},
                        xticklabel style={rotate=45, font=\tiny},
                        ylabel = {\# of bugs},
                        ylabel near ticks,
                        nodes near coords,
                        every node near coord/.append style={color=white, font=\tiny},
                    ]
                    \addplot [fill={rgb,255: red,9; green,99; blue,125}] table [y=fixed, meta=Label, x expr=\coordindex] {\buglocations};
                    \addlegendentry{Fixed}
                    \addplot [fill={rgb,255: red,232; green,189; blue,70}] table [y=confirmed, meta=Label, x expr=\coordindex] {\buglocations};
                    \addlegendentry{Confirmed}
                \end{axis}
            \end{tikzpicture}
        }
        \vspace{-1\baselineskip}
        \caption{
            Bugs categorized by \textbf{locations}.
            CodeGen refers to the code shared by all architectures, so these bugs may affect all architectures.
        }
        \label{fig:bugs:locations}
    \end{subfigure}
    \vfill
    \begin{subfigure}[t]{\linewidth}
        \centering
        \resizebox{\linewidth}{!}{
            \begin{tikzpicture}
                \begin{axis}[
                        width=\linewidth,
                        height=9\baselineskip,
                        ybar stacked,
                        ymin=0,
                        ymax=24,
                        xtick=data,
                        legend style={cells={anchor=west}, legend pos=north east, font=\scriptsize},
                        reverse legend=true, 
                        xticklabels from table={\bugtypes}{Label},
                        xticklabel style={font=\tiny},
                        ylabel = {\# of bugs},
                        ylabel near ticks,
                        nodes near coords,
                        every node near coord/.append style={color=white, font=\tiny},
                    ]
                    \addplot [fill={rgb,255: red,9; green,99; blue,125}] table [y=fixed, meta=Label, x expr=\coordindex] {\bugtypes};
                    \addplot [fill={rgb,255: red,232; green,189; blue,70}] table [y=confirmed, meta=Label, x expr=\coordindex] {\bugtypes};
                \end{axis}
            \end{tikzpicture}
        }
        \caption{
            Bugs categorized by \textbf{causes}.
            Most of the severe bugs are compiler hangs, memory errors, and assertion failures.
        }
        \label{fig:bugs:catogories}
    \end{subfigure}
    \caption{
        Distributions of bugs found by \irfuzzer{}.
        \irfuzzer{} has found \textbf{\numTotalBugs{}} new bugs, out of which \textbf{\numFixedBugs{}} have been fixed.
    }
    \label{fig:bugs}
\end{figure}

\begin{lstlisting}[
  belowskip=-2\baselineskip,
  float,
  language=C++, 
  caption={
      A snippet of code in LLVM where the index of a vector is treated as a signed value.
  }, 
  label=lst:zext,
  basicstyle=\ttfamily\scriptsize,
] 
bool IRTranslator::translateExtractElement(
const User &U, MachineIRBuilder &MIRBuilder) { 
  Register Idx;
  const LLT Ty = LLT::scalar(PreferredVecIdxWidth);
  Idx = MIRBuilder.build(*@\textbf{SExt}@*)OrTrunc(Ty, Idx).getReg(0);
}
\end{lstlisting}

\add{We demonstrate two bugs found by \irfuzzer.}
\sub{\autoref{lst:zext} shows a bug that \irfuzzer found in the LLVM backend.}
\add{\autoref{lst:zext} shows a bug in IR Translator.}
When translating the \ir{} \texttt{extractelement}, the bug extends the index as a signed integer, e.g., translating \texttt{char 255} into \texttt{-1}.
This bug generates incorrect \mir{s} and affects the LLVM backend for seven architectures.
Introduced in LLVM nine years ago, the bug was never noticed for several reasons.
First, it is less common for compiler frontends to generate vector operations, as we have seen in \autoref{sec:eval:csmith}, and is even rarer to use an index that is large enough to wrap around to a negative integer.
However, \irfuzzer{} can generate such a test case easily because it directly mutates on \ir{}.
More importantly, the documentation was ambiguous with respect to the desirable behavior.
The documentation states ``The index may be a variable of any integer type''~\cite{zext} without giving details on how it should be interpreted.
Therefore, when this bug was introduced, it complied with the incomplete documentation at the time.
This exemplifies how complex software interfaces can be incompletely specified, which further justifies our specialized fuzzing.
In this case, we fixed the bug and updated the documentation to reflect the intended interpretation of the index as an unsigned integer.

\begin{figure}[t]
    \centering
    \begin{subfigure}[b]{0.45\linewidth}
        \centering
        \resizebox{\linewidth}{!}{%
            \begin{tikzpicture}[auto]
                \tikzstyle{every node} = [align=center, node distance=1.5cm]
                \node [draw, circle] (BB1)  {BB1};
                \node [draw, circle, right of=BB1] (BB2) {BB2};
                \node [draw, circle, right of=BB2] (BB3) {BB3};

                \path (BB1) edge[cfgedge] node {} (BB2);
                \path (BB2) edge[cfgedge, bend right] node {} (BB3);

                \path (BB2) edge[cfgedge, out=220,in=250,looseness=10] node [below] {\texttt{x == 3}} (BB2);
                \path (BB2) edge[cfgedge, out=290,in=320,looseness=10] node [below] {\texttt{x == 2}} (BB2);
                \path (BB3) edge[cfgedge, bend right] node [above] {\texttt{x == 2}} (BB2);
            \end{tikzpicture}
        }
        \caption{Original CFG}
        \label{fig:hang:a}
    \end{subfigure}
    \hfill
    \begin{subfigure}[b]{0.45\linewidth}
        \centering
        \resizebox{\linewidth}{!}{%
            \begin{tikzpicture}[auto]
                \tikzstyle{every node} = [align=center, node distance=1.5cm]
                \node [draw, circle] (BB1)  {BB1};
                \node [draw, circle, right of=BB1] (BB2) {BB2};
                \node [draw, circle, right of=BB2] (BB3) {BB3};

                \path (BB1) edge[cfgedge] node {} (BB2);
                \path (BB2) edge[cfgedge, bend right] node {} (BB3);

                \path (BB2) edge[cfgedge, loop below] node [below] {\texttt{x == 2 || x == 3}} (BB2);
                \path (BB3) edge[cfgedge, bend right] node [above] {\texttt{x == 2}} (BB2);
            \end{tikzpicture}
        }
        \caption{Optimized CFG}
        \label{fig:hang:b}
    \end{subfigure}
    \caption{\add{
        A piece of code generated by \irfuzzer, simplified to CFG only.
        \texttt{TurnSwitchRangeIntoICmp} transforms \autoref{fig:hang:a} into \autoref{fig:hang:b}, and 
        \texttt{FoldValueComparisonIntoPredecessors} will undo the transformation, causing an infinite loop.
    }}
    \label{fig:hang}
    \vspace{-1\baselineskip}
\end{figure}

\add{

\irfuzzer also found compile hangs.
\autoref{fig:hang} shows a simplified CFG corresponding to the code generated by \irfuzzer{}.
This CFG will cause a compiler hang due to the interaction between two optimization passes.
BB2 in \autoref{fig:hang:a} consists of a \texttt{switch} statement with two self loop edges.
The \texttt{TurnSwitchRangeIntoICmp} optimization attempts to rewrite the condition as a branch predicate because \texttt{x == 2 || x == 3} can be optimized using bit operations, rewriting \autoref{fig:hang:a} into \autoref{fig:hang:b}.
However, the \texttt{FoldValueComparisonIntoPredecessors} optimization converts this code back into a \texttt{switch} statement to reduce the number of comparison operations, turning the CFG back to \autoref{fig:hang:a}.
As a result, a fixed point is never reached, creating an infinite loop.
This bug is hard to trigger since the bug can only be triggered when the \texttt{switch} in \autoref{fig:hang:b} has exactly two destinations (BB2 and BB3), and the switch conditions are consecutive, enabling the \texttt{TurnSwitchRangeIntoICmp} optimization.
This combination is unlikely to be created during manual testing, and can only happen through the interaction of two largely unrelated pieces of code.  Yet, we are able to discover this catastrophic combination through our CFG mutation strategy in a time frame amenable to run fuzzing on every nightly build with little human intervention.
}

We are working closely with the LLVM community to fix the bugs discovered by \irfuzzer.
\numFixedBugs{} bugs were fixed, \backPortFix{} of which were back ported to LLVM 15 as security patches.
\sub{
Despite heavy testing, \textbf{all} fixed bugs are introduced before LLVM 15.
}
\add{
The developers confirmed that all the bugs that we reported and they fixed had been introduced prior to LLVM 15. 
Despite heaving testing, they remained in LLVM 15 until \irfuzzer discovered them.
}
This demonstrates that specialized fuzzing for compiler backend is necessary, and it provides actionable insight to developers.

\add{
    \subsubsection{Accuracy}
    \label{sec:eval:types:accuracy}
    \irfuzzer{} guarantees to generate valid LLVM IR.
    Since IR is the input to backends, a robust backend should take any valid IR without crashing, so any crash indicates either a bug or an incomplete feature in the backend.
    \irfuzzer{} found \numTotalBugs{} bugs, all of which we confirmed and reported to the developers. Of these bugs, 
    \numFixedBugs{} have been fixed, and \backPortFix{} bugs were back-ported as security patches.
    This shows that the developers agree that these are true bugs regardless of whether C programs corresponding to the IR exist.
}

We can answer \textbf{RQ4} now.
\sub{
    We found \numHangs{} compiler hangs, \numMemory{} memory errors, and \numAssertionErrors{} assertion failures.
    We also found \numLogicErrors{} logic errors and \numMissingPatterns{} missing patterns in the matcher table.
    In total, we found \numTotalBugs{} new bugs, out of which \numFixedBugs{} have been fixed and \backPortFix{} have been back ported to LLVM 15 as security patches.
}
\add{
    In total, \irfuzzer found \numTotalBugs{} new bugs. All have been confirmed, \numFixedBugs{} have been fixed, and \backPortFix{} have been back ported to LLVM 15 as security patches.
    All these bugs were found only by \irfuzzer{} and not by any of the baseline fuzzers.
    These bugs contain \numHangs{} compiler hangs, \numMemory{} memory errors, and \numAssertionErrors{} assertion failures.
    We also found \numLogicErrors{} logic errors and \numMissingPatterns{} missing patterns in the matcher table.
}

\section{Related work}
\label{sec:related}

Prior work has focused on compiler testing~\cite{10.1145/3363562, ma2023survey, 10.1145/3360581}.
One popular approach is to generate inputs for compilers to compile.
Purdom\cite{purdom1972sentence} generates program based on context free grammar.
Superion\cite{8811923} and Nautilus\cite{aschermann2019nautilus} also relies on context free grammar for fuzzing.
However, context free grammar based methods cannot generate semantically meaningful programs.
These efforts are effective in testing frontend parsers, but cannot reach the backends effectively.

While many fuzzers are testing the frontend of the compiler using grammar based method~\cite{zeller2019fuzzing}, some work also tests the correctness of middle-end~\cite{lopes2021alive2, mansky2010framework, 10.1145/3591295, 10.1145/3166064}.
To the best of our knowledge, \irfuzzer{} is the first one to verify the compiler backend using an architecture independent approach.

Some work does end-to-end tests using high-level programming languages.
CSmith~\cite{CSmith}, YARPGen~\cite{10.1145/3428264}, and Grayc~\cite{GreyC} generate C and C++ programs.
AI has also been used for program generate for the purpose of compiler testing ~\cite{DeepFuzz, deng2023large, xia2023universal}.
However, end-to-end testing implies that there is a need to create a generator for every language, like JavaScript~\cite{9152648}, Rust~\cite{rust}, and Java~\cite{java1, java2, JITFuzz}.
POLYGLOT\cite{POLYGLOT} introduced a language-free IR and mutator based on it.
Most fuzzers have no feedback from the compiler.
Even though Grayc~\cite{GreyC} introduced branch coverage feedback, it was unable to trigger backend bugs due to language limitations and compiler optimizations.
Instead of directly generating a program, Equivalence Modulo Inputs~\cite{emi, 10.1145/2814270.2814319} mutates an existing C program to preserve its semantics.
Therefore, the program before and after mutation should have the same behavior.
Combining CSmith and EMI, Lidbury et al. mutate program to test OpenCL compiler ~\cite{Lidbury}.
However, the language limits these work, since the generator cannot help when the language frontend cannot exercise a feature in the compiler.

Formal verification is another valuable part of compiler verification~\cite{10.1145/966221.966235}.
Verasco~\cite{Jourdan-LBLP-2015} is a formally verified C analyzer.
CompCert~\cite{10.1145/1538788.1538814} is a compiler for a subset of C that is formally verified.
There is work that verifies other languages, like Rust~\cite{10.1145/3360573} and Lustre~\cite{Bourke-BDLPR-2017}.
However, formal verification cannot scale to large compilers like LLVM, therefore it has a limited impact in the community.

There is also work that considered generating a valid intermediate representation for testing purposes.
\fuzzmutate{} directly generates LLVM IR~\cite{FuzzMutate}.
However, FuzzMutate has no feedback unless combined with fuzzers like AFL++~\cite{AFLplusplus} or libFuzzer~\cite{serebryany2016continuous}.
Some work focus on testing of a specific compiler~\cite{10.1145/3597926.3598053, 10.1145/3527317}.
Tzer focuses on IR mutation in the context of a tensor compiler~\cite{10.1145/3527317}.
However, Tzer relies on LLVM's Coverage Sanitizer that only tracks code coverage.
Similar to \irfuzzer{}'s approach, ClassMing directly mutates on Java byte code~\cite{8811957}.
Neither Tzer nor ClassMing designed a feedback approach, except for branch coverage.
However, as we demonstrate in ~\autoref{sec:eval:matcher}, a customized feedback metric can greatly help the fuzzer to reach deeper into the code base.
With the development of large language models (LLM), it has been used more and more in fuzzing and code generation \cite{zhang2024llamafuzzlargelanguagemodel,10.1145/3597503.3639121,yang2023whitefox, 10.1145/3650212.3680342}. 
However, LLM doesn't guarantee the correctness of input like \irfuzzer{} does.

\section{Conclusion}
\label{sec:conclusion}

We described \irfuzzer{}, a fuzzer specializing in fuzzing LLVM instruction selection.
To generate semantically and syntactically correct inputs, we identified the challenges in IR generation that did not exist in high-level language generation.
We created a mutator that maintained semantic correctness by splitting blocks and inserting a \sCFG in between.
Then, we ensured that the \ir{s} that we inserted were syntactically correct using a descriptive language to model all \ir{s}.
Therefore, the IR program that \irfuzzer generated could always be compiled by the backend.
We proposed a new metric to track the coverage of the matcher table and decoded the coverage table to guide mutation.

Our evaluation shows that \irfuzzer{} outperformed existing backend and end-to-end \sota fuzzers.
\irfuzzer{} achieved higher matcher table coverage on all the LLVM backend architectures.
\irfuzzer{} is also efficient enough to become part of the development process.

\irfuzzer{} identified \numTotalBugs{} new, confirmed bugs in upstream LLVM code. Upon receiving our bug report, the developers have fixed 
\numFixedBugs{} bugs and back-ported \backPortFix{} fixes to LLVM 15. This demonstrates that \irfuzzer is effective in finding bugs in LLVM backend and provides useful, actionable insights to LLVM developers. Our experience shows that there are fertile opportunities for specialized fuzzing despite popular end-to-end compiler testing.
\section*{Acknowlegement}

This work is partially supported by UC Noyce Initiative.

\printbibliography
\newpage

\end{document}